\documentclass[usenatbib]{mn2e}
\usepackage{apjfonts}
\usepackage{epsfig}
\usepackage{amssymb}
\usepackage{amsmath}
\usepackage{fixltx2e} 

\newcommand{\mdot}{\dot{m}}

\newcommand{\msun}{M_{\sun}}

\newcommand{\plotside}[1]
 {\centering \leavevmode \includegraphics[width={0.95\textwidth}]{#1}}
\newcommand{\acknowledgments}{\begin{small}\section*{Acknowledgments}\end{small}}
\newcommand\altaffilmark[1]{$^{#1}$}
\newcommand\altaffiltext[1]{$^{#1}$}
\title[Multi-Stage AGN Feedback]{Quasar Feedback: More Bang for Your Buck}

\author[Hopkins \&\ Elvis]{
\parbox[t]{\textwidth}{ 
Philip F. Hopkins\altaffilmark{1}\thanks{E-mail:phopkins@astro.berkeley.edu},
\&\ Martin Elvis\altaffilmark{2} } 
\vspace*{6pt} \\
\altaffiltext{1}{Department of Astronomy, University of California
  Berkeley, Berkeley, CA 94720} \\
\altaffiltext{2}{Harvard-Smithsonian Center for Astrophysics, 60
  Garden Street, Cambridge, MA 02138} \\
}

\date{Submitted to MNRAS, March 29, 2009}
\begin{document}
\maketitle
\label{firstpage}

\begin{abstract}
We propose a ``two-stage'' model for the effects of feedback from a bright quasar 
on the cold gas in a galaxy. It is difficult for winds or 
other forms of feedback from near the accretion disk to directly 
impact (let alone blow out of the galaxy) dense molecular clouds at $\sim$kpc. But 
if such feedback can drive a weak wind or outflow in the hot, diffuse ISM 
(a relatively ``easy'' task), then in the wake of such an outflow passing over a 
cold cloud, a combination of instabilities and simple pressure gradients 
will drive the cloud material to effectively expand in the direction perpendicular 
to the incident outflow. This shredding/expansion (and the corresponding decrease in 
density) may alone be enough to substantially suppress star formation in the host. 
Moreover, such expansion, by even a relatively small factor, 
dramatically increases the effective cross section of the cloud material and makes it 
much more susceptible to both ionization and momentum coupling from absorption of 
the incident quasar radiation field. We show that even a moderate effect of this nature 
can dramatically alter the ability of clouds at large radii to be fully ionized 
and driven into a secondary outflow by radiation pressure. Since the 
amount of momentum and volume which can be ionized by observed quasar radiation 
field is more than sufficient to affect the entire cold gas supply once 
it has been altered in this manner (and 
the ``initial'' feedback need only initiate a moderate wind in the low-density 
hot gas), this reduces by an order of magnitude the required energy budget 
for feedback to affect a host galaxy. 
Instead of $\sim5\%$ of the radiated energy ($\sim100\%$ momentum) 
needed if the initial feedback must directly heat or ``blow out'' the galactic gas, 
if only $\sim0.5\%$ of the luminosity ($\sim10\%$ momentum) 
can couple to drive the initial hot outflow, this 
mechanism could be efficient. 
This amounts to hot gas outflow rates from near the 
accretion disk of only $\sim5-10\%$ of the BH accretion rate.
\end{abstract}


\section{Introduction}
\label{sec:intro}

Observations have established that the masses of supermassive black 
holes (BHs) are tightly correlated with various host 
galaxy properties \citep{magorrian,FM00,Gebhardt00,
hopkins:bhfp.obs,aller:mbh.esph}. Together with constraints indicating that 
most of the BH mass is assembled in optically bright quasar\footnote{In 
this paper, we use the term ``quasar'' loosely as a proxy for 
high-Eddington ratio activity, rather than as a reference to 
specific optical properties.} phases 
\citep{Soltan82,salucci:bhmf,yutremaine:bhmf,hopkins:old.age}, 
this has led to the development of models where feedback processes 
from accretion self-regulated BH growth at a critical mass 
\citep{silkrees:msigma,dimatteo:msigma,murray:momentum.winds}. 
Gas inflows triggered by some process fuel rapid BH growth, until 
feedback begins to expel nearby gas and dust. This ``blowout'' 
results in a short-lived, bright optical quasar that, having expelled its 
fuel supply, fades and leaves a remnant on the observed 
BH-host correlations \citep{hopkins:lifetimes.methods,hopkins:lifetimes.obscuration}. 
These scenarios have been able to explain many quasar observables, 
including luminosity functions, lifetimes, and BH mass functions 
\citep{hopkins:lifetimes.interp,hopkins:merger.lfs,
hopkins:groups.qso,hopkins:seyfert.bimodality,volonteri:xray.counts,
menci:sam,somerville:new.sam,lapi:qlf.sam,
tortora:2009.agn.jet.fb.and.ell.colors}. 

It is much less clear, however, what the impact of whatever feedback 
regulates BH growth will be on the host galaxy. In models, such feedback 
is invoked to explain the rapid ``quenching'' of star formation and sustained lack of 
cooling in massive galaxies 
\citep{granato:sam,scannapieco:sam,croton:sam,hopkins:groups.ell,
antonuccio-delogu:2008.jet.fb.destroying.sf.clouds}. 
The argument in the models is that, under various simple assumptions, 
if sufficient energy or momentum is injected into the ISM near the BH on a timescale 
short enough to halt accretion, then it will yield a supersonic 
pressure or momentum-driven outflow that propagates to large scales 
\citep[see e.g.][]{monaco:feedback,hopkins:qso.all,
shin:mech.agn.fb.constraints}. 

But the actual mechanisms of feedback and physics of the ISM rlevant for this 
remain highly uncertain. 
Highly energetic outflows are associated 
with bright quasars \citep[for a review, see][]{veilleux:winds}; these range 
from intense winds ($v\sim10^{4}\,{\rm km\,s^{-1}}$) 
associated with the central engine seen in the broad 
emission line regions and broad absorption line quasars \citep[e.g.][]{weymann:BALs} 
to more moderate outflows ($v\sim10^{2}-10^{3}\,{\rm km\,s^{-1}}$) associated 
with the narrow line region and the ``warm absorber''
\citep{laor:warm.absorber,crenshaw:nlr} 
as well as with small-scale quasar absorption and occultation systems 
\citep[e.g.][]{mckernan:1998.agn.occultation.by.clumpy.outflow,
turner:2008.clumpy.agn.disk.wind,miller:2008.clumpy.agn.disk.wind}. 
Indeed, high-velocity winds 
driven near the accretion disk are theoretically hard to avoid 
\citep[see e.g.][]{blandfordpayne:mhd.jets,begelman:agn.compton.heating,
koniglkartje:disk.winds,elvis:outflow.model,
proga:disk.winds.2000,proga:disk.winds}. 
However, these are probably tenuous, with an initial mass-loading 
$\lesssim \dot{M}_{\rm BH}$
\citep[although in at least some cases, these outflows are extremely 
dense, and might have much higher mass-loading factors; see e.g.][]{
hall:2003.high.density.balqso.outflows,hall:2007.agn.outflow.substructure}. 
It is not clear whether such ``hot'' outflows could efficiently entrain gas 
at larger radii. If most of the gas mass of the galaxy is at some appreciable 
fraction of the galaxy effective radius $R_{e}$ 
and in the form of cold, dense giant molecular clouds (GMCs), 
then it is difficult to imagine such a diffuse wind directly ``launching'' 
the clouds out of the galaxy. It remains unclear whether, in fact, the momentum 
associated with the winds that {\em are} known to emanate from the 
central engine of a quasar is sufficient to unbind the cold 
gas in the host \citep[see e.g.][]{baum:radio.outflows,dekool:large.outflow.1,
dekool:large.outflow.2,Steenbrugge:outflow.mdot,
holt:merger.radio.warm.outflow,gabel:large.outflow,
krongold:warm.absorber.outflow.rate,
krongold:seyfert.outflow,batcheldor:outflow.mechanism,tremonti:postsb.outflows,
McKernan:agn.fb.outflow.rates, ganguly:qso.outflows,
prochaska:qso.outflow}. 

In this paper, we argue that it is not necessary that the small-scale, high-velocity 
AGN outflows directly entrain any cold gas at scales $\sim R_{e}$. 
Rather, so long as these are sufficient to drive a significant wind 
in the ``hot'' diffuse ISM, then clouds will be effectively destroyed or deformed and 
``secondary'' feedback mechanisms -- namely the radiative effects of 
dust absorption and ionization -- will be able to 
act efficiently on the cold gas at large scales. This will effectively terminate star formation on a 
short timescale, with greatly reduced energy/momentum requirements for 
the ``initial'' outflow drivers.

\section{Radiative Feedback In the Presence of Hot Outflows}
\label{sec:radiative}

\begin{figure*}
    \centering
    \plotside{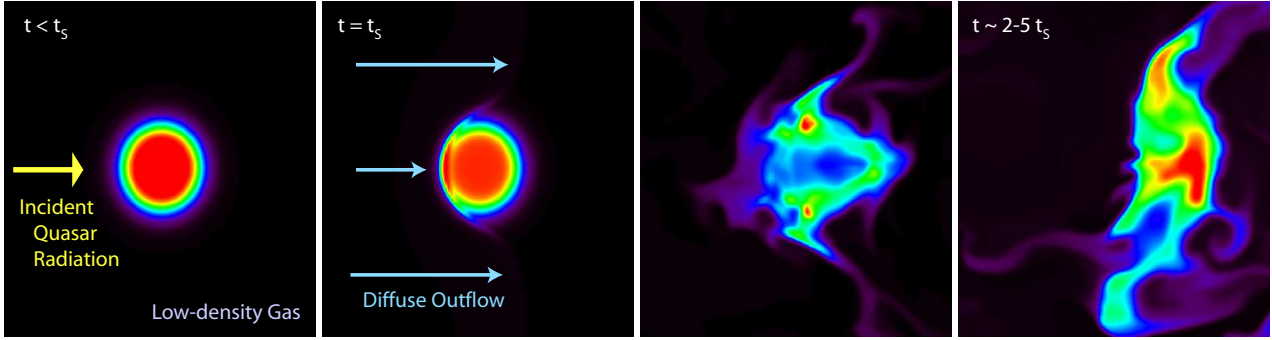}
    \caption{An illustration from a simple simulation of the effects of a hot outflow on 
    radiation feedback. {\em Left:} Initial dense cloud in pressure 
    equilibrium with the diffuse ISM. Ionization and momentum flux from 
    the quasar is negligible, because the effective surface area for absorption is 
    small. 
    {\em Center:} At $t=t_{s}$, an outflow generated in the hot/diffuse ISM 
    hits the cloud. Ambient pressure drops in the wake (low-pressure regions form 
    on the perpendicular sides of the cloud). A mix of instabilities shortly cause the 
    cloud to mix in the perpendicular direction. 
    {\em Right:} After a few cloud crossing times, the mixing 
    has increased the effective cross section by a factor 
    $(R_{c}/R_{0})^{2}\gtrsim10$ in the perpendicular direction. 
    The cloud may now (with lower density and higher cross section to 
    absorb quasar photon momentum) be vulnerable to radiative 
    feedback, and will be accelerated to $v\sim v_{\rm esc}$. 
    \label{fig:cartoon}}
\end{figure*}

Consider a typical galaxy, where the ISM gas is composed of a 
mix of diffuse warm/hot gas and cold clouds.\footnote{By 
``diffuse'' ISM, we refer to the warm/hot phases of the ISM, 
with effective temperatures $T\sim 10^{4}-10^{6}\,$K and 
densities $n_{H}\sim 10^{-2}-10^{0}\,{\rm cm^{-3}}$. The ``clouds,'' 
on the other hand, represent the cold phase (but {\em not} 
collapsing molecular cloud cores) with 
$T\lesssim 100\,$K molecular gas and 
$n_{H}\sim 10^{2}-10^{4}\,{\rm cm^{-3}}$. The volume filling 
factor of clouds is small, but by mass, they dominate 
the total gas mass; the hot gas with filling factor near 
unity has a typical mass fraction $f_{\rm hot}\lesssim0.1$ 
\citep[see e.g.][]{mckee.ostriker:ism}.} 
Radiation will always act on the cold clouds (in the form 
of ionization and momentum injection from absorbed photons), 
but they may be too dense and self-shielding to be 
significantly affected. 
In such a case, one could invoke a blastwave or cold 
shell, driven by AGN feedback on small scales, to 
entrain this material. Various models and 
simulations have shown that if feedback needs to directly launch 
a blastwave in {\em both} the hot and cold gas together (sufficient 
to entrain most of the galactic gas), 
then an efficiency $\eta_{\rm E}\sim0.05$ is the relevant value 
(for energy injection where the feedback $\dot{E} = \eta_{\rm E}\,L = 
\eta_{\rm E}\,\epsilon_{r}\,\dot{M}_{\rm BH}\,c^{2}$; $\epsilon_{r}\sim0.1$ 
is the radiative efficiency). If the outflow is instead 
momentum-driven, the relevant value for driving hot+cold phases 
is $\eta_{\rm p}\sim1$ ($\dot{p} = \eta_{\rm p}\,L/c$). 

If, however, the ``initial'' feedback from the central source 
need only drive a wind in the low-density hot gas, and 
does not necessarily directly entrain the cold clouds, 
then the energy required will be much less. 
In both cases above, the efficiency needed 
simply scales linearly with the mass of material to be driven 
(scaling with its binding energy or momentum, respectively). 
So, if a fraction $f_{\rm hot}$ of the gas is in the hot, diffuse ISM, 
and only that needs to be initially driven, we obtain
\begin{equation}
\eta_{\rm E} = \frac{\dot{E}_{\rm fb}}{L} \sim 0.05\,f_{\rm hot}
\end{equation}
for energy-driven and 
\begin{equation}
\eta_{\rm p} = \frac{\dot{p}_{\rm fb}}{L/c} \sim f_{\rm hot}
\end{equation}
for momentum-driven outflows. 
For typical $f_{\rm hot}\lesssim0.1$ \citep[e.g.][]{blitz:gmc.properties}, 
this implies an order-of-magnitude 
reduction in the necessary feedback input for ``interesting'' 
behavior. And in detail, since the hot-phase gas is already virialized 
(rather than in e.g.\ a cold disk), the efficiency gains may be even higher. 
The question is then, if most of the mass is in cold clouds and only 
the hot gas is affected by this initial outflow, will any interesting 
behavior result?

Consider a cold molecular cloud in a galaxy. The cloud has a mass 
$M_{c}$, and an initial (``equilibrium'') effective spherical radius $R_{0}$. 
In the observed typical ISM, these are related by 
\begin{equation}
M_{c}\sim300\,\msun\,R_{0,{\rm pc}}^{2}
\label{eqn:cloud.sizemass}
\end{equation} 
\citep[where $R_{0,{\rm pc}}\equiv R_{0}/{1\,{\rm pc}}$;][]{blitz:h2.pressure.corr}. 
Since we will later allow the cloud to stretch or deform, define 
the instantaneous radius of the cloud as $R_{c}$. 
The cloud resides at a spherical distance $r$ from the center of 
a \citet{hernquist:profile} profile bulge of total mass 
$M_{\rm bul}$ and characteristic 
scale length $R_{\rm eff}$, which hosts a BH on the characteristic BH-host 
relations, $M_{\rm BH}=\mu_{\rm BH}\,M_{\rm bul}$ 
\citep[$\mu_{\rm BH}\approx0.0014$;][]{haringrix}. 
The BH radiates with a luminosity 
\begin{equation}
L_{\rm QSO}=\mdot\,L_{\rm Edd}\ ,
\end{equation} 
where $\mdot$ is the dimensionless Eddington ratio and 
$L_{\rm Edd}=1.3\,M_{\rm BH,8}\times10^{46}\,{\rm erg\,s^{-1}}$ 
is the Eddington luminosity 
($M_{\rm BH,8}=M_{\rm BH}/10^{8}\,\msun$). 

We are interested in the case where the cold clouds are 
embedded in some kind of hot outflow generated by the ``primary'' AGN feedback. 
Specifically, assume that the quasar somehow succeeds in driving an 
outflow through the diffuse warm/hot ISM: to be conservative, the outflow 
can be tenuous and we will assume that the outflow 
``impacting'' the cloud carries negligible momentum 
compared to the binding momentum of the cloud. In other words, assume 
that at some smaller scale, a wind or outflow is generated sufficient to 
sweep up the tenuous, diffuse ISM at large radii, but insufficient to affect cold dense 
clouds, which contain most of the ISM gas mass. 

This problem of the survival of cold clouds in a post-shock hot medium is 
well-studied in the context of star formation and supernova feedback
\citep[see e.g.][and references therein]{kmc:shock.cloud}. 
In general, the collision of a shock or 
wind with velocity $v_{s}$ with a cloud of initial characteristic 
(quasi-spherical) radius $R_{0}$ and density contrast $\chi$ (ratio of 
cloud density to external medium density $\chi\equiv n_{c}/n_{0}$) 
will launch secondary shocks within the cloud with velocity 
$v_{s}/\chi^{1/2}$. This defines a ``cloud crushing'' timescale 
$t_{\rm cc} = \chi^{1/2}\,R_{0}/v_{s}$.
In the simple case of a 
pure hydrodynamic strong shock, if $t_{\rm cc}$ is 
much less than the characteristic timescales for the density to 
change behind the shock and $v_{s}/\chi^{1/2}$ is 
comparable to or larger than the characteristic internal velocities 
of the cloud, then the cloud will be stretched and ``shredded'' 
by a combination of Rayleigh-Taylor and Kelvin-Helmholtz instabilities 
on a timescale $\sim$a few $t_{\rm cc}$ \citep[see e.g.][]{kmc:shock.cloud,
xustone:cloud.shock.3d,fragile:shock.cloud.2004,
orlando:cloud.shock.w.cooling,nakamura:varied.cloud.shock.interactions}. 
Given the definitions above, for $v_{s}\sim v_{\rm esc}$, 
$t_{\rm cc}$ is much less than the dynamical timescales of interest 
at all the spatial scales of interest ($t_{\rm cc}\ll 10^{7}\,{\rm yr}$ for 
all $r\gtrsim\chi^{1/2}\,R_{0}$ -- i.e.\ for clouds not in the very nuclear regions). 

The material being mixed off of the surface of the cloud from 
these instabilities expands into 
and mixes with the low-pressure zones created by the passage of the 
shock on the sides of the cloud. This leads to an effective net expansion of 
the cloud by a factor $\sim\chi^{1/2}$ in radius in the perpendicular shock direction. 
Eventually, despite the initial compression, reflection shocks lead to a 
expansion by a factor $\sim2$ in the parallel shock direction, bringing the 
original cloud material into an effective density and pressure equilibrium with the 
external medium. The surface area of the cloud can increase dramatically; 
for our purposes, we are interested in the effective 
cross section the cloud presents to the perpendicular shock direction. 
The radii defined above should be thought of in this manner: 
the initial cloud cross section to the hot shock is $\pi\,R_{0}^{2}$; 
post-shock, the effective cross section owing to this expansion and 
equilibration is $\pi\,R_{c}^{2}\sim\chi\pi\,R_{0}^{2}$. 

We illustrate this behavior with a simple toy model system in 
Figure~\ref{fig:cartoon}. Specifically, we show an example of a hydrodynamic simulation 
of an idealized system, using the {\textsc{ZEUS}} code \citep{zeus:a,zeus:b}. The initial 
conditions consist of a Plummer sphere cloud embedded in a uniform background, with density 
contrast of $\chi=100$ (peak density of the cloud relative to background), 
with the initial system in pressure equilibrium (uniform pressure), 
and periodic boundary conditions in a large grid. At time $t=t_{S}$ 
the low-density material is rapidly accelerated into a mach $\sim2$ wind. 
Color encodes the gas density, from black (the arbitrary background density) to 
red (the initial maximum). 
Note that the example shown is purely for illustrative purposes -- 
we do not include many possible complexities, such as gas cooling, 
star formation, or magnetic fields. The behavior of clouds in response to 
outflows with such sophistications has been extensively studied in the 
references above, and more detailed extensions of such simulations to 
the regime of interest here will be the subject of future work. Nevertheless, 
this simple experiment illustrates much of the important qualitative behavior. 

The qualitative behavior we care about -- the mixing/stretching/deformation 
of the cloud in the perpendicular shock direction leading to an increase in the 
effective cross section of the cloud -- is in fact quite general. 
Simulations have shown that the same instabilities operate regardless of 
whether the ``hot outflow'' is a strong shock, weak shock, or wind 
(since we assume the hot material is being unbound in this wind, it cannot be substantially 
sub-sonic). The timescale of 
cloud expansion increases by a factor of a few in the weaker wind case, but it is still much less 
than the relevant local galactic dynamical times 
\citep{kmc:shock.cloud,jones:mhd.supersonic.cloud}. 
The process is also similar in the case of a cloud being impacted by 
AGN jets, despite the different densities, temperatures, and magnetic field 
states associated with jets and ``bubbles'' \citep[see][]{krause:2007.turbulent.jet.cocoons,
antonuccio-delogu:2008.jet.fb.destroying.sf.clouds}. 
A cloud could in principle be stabilized against such instabilities 
by being strongly magnetically dominated 
\citep{maclow:mhd.shock.cloud,
junjones:shock.cloud.mhd.cosmic.rays,fragile:shock.cloud.2005,
shinstone:3d.mhd.shock.cloud}. However, in this limit, as the hot outflow 
sweeps up material, the pressure of the diffuse ISM trailing the outflow 
will decline as a steep function of time $t/t_{\rm cc}$ 
\citep[][]{ostrikermckee:blastwaves}. Since, in this limit, the cloud 
is then over-pressurized, it will expand isothermally as the exterior 
post-shock pressure drops (the free expansion/equilibration time of the cloud 
being short compared to the other timescales of interest). 
Because this stops when the system is equilibrated, the 
``effective'' net expansion of $R_{c}$ is the same as in the hydrodynamic 
shock case, even though the details are quite different. 

If nothing more were to happen to the cloud, this would only suppress 
star formation for a short time. The cooling instabilities that 
produced the cloud in the first place would 
operate. In the ``typical'' ISM, clouds mix in the wake of 
stellar or supernovae-driven outflows until they reach equilibrium 
with the ISM and 
re-cool into new clouds. 

However, we are interested in all of this occuring in the background of a luminous 
AGN, which will both ionize and exert a radiation pressure force. 
The cloud -- especially a realistic cloud with a large dust mass 
and corresponding opacity -- is optically thick\footnote{
For any incident spectrum where a 
significant fraction of the incident 
energy is in the optical/UV or higher wavelengths, the effective optical depth 
from dust within the cloud will be 
$\tau\sim1$ \citep{murray:momentum.winds,thompson:rad.pressure} -- 
this is simply a statement that the 
clouds will be optically thick to some portion of that SED (whether 
ionizing photons in the far UV, or, if the cloud is not yet ionized, 
then there is dust which will absorb in the optical and IR) and re-radiate 
that energy. The cloud can effectively be thought of as a single absorbing ``mega-grain'' 
with effective cross section $\sim\pi R_{c}^{2}$.
} 
to the quasar radiation, with an 
effective cross-section $\Delta\Omega\sim(\pi\,R_{c}^{2})/(4\pi\,r^{2})$. 
There is therefore an inescapable deposition of 
photon momentum from the radiation field with a 
deposition rate $\dot{p}_{\rm rad} = L_{\rm abs}/c$ 
where $L_{\rm abs}=L_{\rm QSO}\,\Delta\Omega$. 
Comparing this to the gravitational force 
$F_{\rm grav} = -M_{c}\partial \phi /\partial r$ defines 
an effective Eddington limit for the cloud: if the absorbed flux exceeds some limit 
the cloud will be unbound (equivalently, the absorbed momentum 
in a single dynamical time, over which the cloud could redistribute 
that momentum, will exceed the cloud binding momentum 
$\sim M_{c}\,v_{\rm esc}$). 
This limit is when the two are equal: 
equivalently 
\begin{equation}
\frac{L_{\rm QSO}}{c}\,
\frac{\pi\,R_{c}^{2}}{4\pi\,r^{2}} = M_{c}\,\frac{G\,M_{\rm bul}}{(r+R_{\rm eff})^{2}}\ .
\label{eqn:prad.vs.pedd}
\end{equation}

Assuming that the cloud lies on the observed size-mass 
relation (Equation~\ref{eqn:cloud.sizemass}) 
and that the galaxy lies on the $M_{\rm BH}-M_{\rm bul}$ relation, 
this reduces to the criterion for unbinding the cloud: 
\begin{equation}
{\Bigl(}\frac{R_{c}}{R_{0}}{\Bigr)} \gtrsim 2.7\ \mdot^{-1/2}\,{\Bigl (} \frac{r}{r+R_{\rm eff}}  {\Bigr )}\ . 
\label{eqn:rcrit.prad}
\end{equation}
In other words, a cloud on the ``normal'' size mass relation at 
large radii $r\sim R_{\rm eff}$ is 
sufficiently dense and sufficiently high column-density to avoid being 
unbound by radiation pressure. But if the effective size of the cloud 
(the effective coupling surface area) could be increased by a factor of a 
couple, or the effective column lowered, the cloud would rapidly be unbound 
by the incident radiation field momentum. This condition is easily 
satisfied in post-shock clouds. 

Figure~\ref{fig:cartoon} also includes and illustrates this effect. 
Specifically, we include a very simple, time-independent momentum 
deposition rate in the ``facing'' cells to the incident radiation field, 
which we simply approximate as all cells with a density 
above $\gtrsim3\,$ times the initial background density but with 
no cell at $x<x_{i}$ above this density (i.e.\ implicitly assuming that 
such a cell would shield the cells ``behind'' it with respect to the incident radiation). 
The magnitude of this is initialized such that at $t<t_{S}$ the 
``total'' deposition rate over the surface of the cloud is equal to a small 
fraction of the ``binding'' momentum over the dynamical time 
(assuming $v_{s}\sim V_{c}$), $\sim0.01\,M_{c}\,v_{s}/(R_{c}/v_{s})$. But the 
details make little qualitative difference to the global acceleration. 

A similar effect pertains to the ionization of the cloud 
(although this is not explicitly included in Figure~\ref{fig:cartoon}). 
Ignoring geometric effects of photon diffusion, the 
volume of a cloud of mean density $n_{c}$ ionized 
is $V_{\rm ion}=\dot{N}/n_{c}^{2}\,\beta$, where 
$\beta\approx2\times10^{-13}\,{\rm cm^{3}\,s^{-1}}$ is the 
recombination coefficient for gas at the temperature for 
hydrogen ionization ($T_{e}\sim10^{4}\,$K) and $\dot{N}$ 
is the rate at which ionizing photons hit the cloud. 
The total rate of production of ionizing photons from the quasar is 
$\dot{N}_{Q}=\lambda L/ h\nu_{912}$ \citep[$\lambda\approx0.07$ 
comes from a proper integration over the quasar spectrum; here from][]{hopkins:bol.qlf}, 
and a fraction $\Delta\Omega$ are incident on the cloud. 
Together with the typical values above, this 
implies that clouds will be ionized to a depth $h_{\rm ionized}$
\begin{equation}
\frac{h_{\rm ionized}}{R_{c}}\approx 10^{-3}\,\mdot\,R_{0,\rm pc}\,
{\Bigl (}\frac{r+R_{\rm eff}}{r}{\Bigr )}^{2}\,
{\Bigl (}\frac{R_{c}}{R_{0}}{\Bigr )}^{5}.
\label{eqn:hionized}
\end{equation}
Give the cloud size-mass relation, this is equivalent to the statement that 
all clouds below a mass 
$M_{c}\lesssim10^{8}\,\msun\,(R_{c}/R_{0})^{-10}$ will be self-shielded 
at $r\gtrsim R_{\rm eff}$. 
For typical clouds, the depth ionized is clearly quite small; but there is a steep 
dependence on cloud radius. As $R_{c}$ increases, 
the ionized depth increases by a factor $\propto R_{c}^{2}$ owing to the 
increased photon capture cross-section and a factor 
$\propto R_{c}^{3}$ owing to the decreased density lowering the 
recombination rate.

\section{Implications in a Global Feedback Scenario}
\label{sec:global}

There are many caveats to the simplified derivations above: 
clouds have some size and mass spectrum, and are distributed at various radii, 
with the background quasar changing in time. 
Nevertheless, embedding this in 
more detailed models for AGN feedback, the results are interesting. 

Consider an $\sim L_{\ast}$ bulge with 
$M_{\rm bul}=10^{11}\,\msun$ and $M_{\rm BH}=1.4\times10^{8}\,\msun$, 
with a \citet{hernquist:profile} density profile 
and scale radius $R_{\rm eff}=4\,$kpc. Assume gas traces stars (initially), 
with mass fraction $f_{\rm gas}=0.1$, and that $90\%$ of the gas is in 
cold clouds while $10\%$ is in a hot diffuse phase (which we assume 
is in hydrostatic equilibrium). 
This yields a density profile of hot or cold gas of 
\begin{equation}
\rho_{i} = f_{i}\,f_{\rm gas}\,\frac{M_{\rm bul}}{2\pi}\,\frac{R_{e}}{R\,(R+R_{e})^{3}}
\end{equation}
where $f_{i}$ represents the fraction in either the hot or the cold phase, 
($f_{\rm hot}=0.1$, $f_{\rm cold}=0.9$). 

At a given radius (within some small radial annulus), the cold gas 
(with mean volume density given above) 
is assumed to be locked 
into cold clouds with a small volume filling factor. The clouds 
are initially placed on the observed size-mass relation (Equation~\ref{eqn:cloud.sizemass}; 
determining an initial radius $R_{0}$ for each cloud of mass $M_{c}$), 
and are distributed in mass according to the observed mass spectrum  
${\rm d}N/{\rm d}M_{c}\propto M_{c}^{-1.8}$ \citep[up to 
a maximum $M_{c}=10^{7}\,\msun$; see][
and references therein]{rosolowsky:gmc.mass.spectrum}. Given 
some total cold gas mass $\rho_{\rm cold}$ per unit volume 
in an annulus, this mass spectrum, 
integrated from arbitrarily small cloud mass (it makes no difference 
if we adopt some lower-mass cutoff) to the maximum $M_{c}$, 
must integrate to $\rho_{\rm cold}$ per unit density; i.e.\ this determines 
the number density $n_{c}$ of clouds at each galaxy 
radius $R$ and mass interval $M_{c}\rightarrow M_{c}+{\rm d}M_{c}$ (and corresponding 
initial cloud radius $R_{0}(M_{c})$). 

At a time $t=0$, we assume that the BH ``turns on,'' radiating (initially) 
at the Eddington limit. 
We allow it to drive a shock/wind through the diffuse ISM 
according to the analytic solutions derived 
in \citet{hopkins:seyferts}. In these models, the AGN is assumed to drive 
a simple, Sedov-Taylor-type outflow via any ``small-scale'' feedback 
channel. The detailed behavior is derived 
and compared with hydrodynamic simulations 
in \citet{hopkins:seyferts}, but can be simply summarized as follows: 
the BH, once on, couples a fraction $\eta\approx0.05\,f_{\rm hot}$ 
of its luminous energy to the diffuse-phase gas in its vicinity. 
Because the spatial and timescales in the vicinity of the BH are 
small compared to the rest of the galaxy, this appears to the 
galaxy as a point-like energy injection in a hot medium. The result 
is therefore a roughly self-similar (power-law) Sedov-Taylor-type outflow. 

\begin{figure*}
    \centering
    \plotside{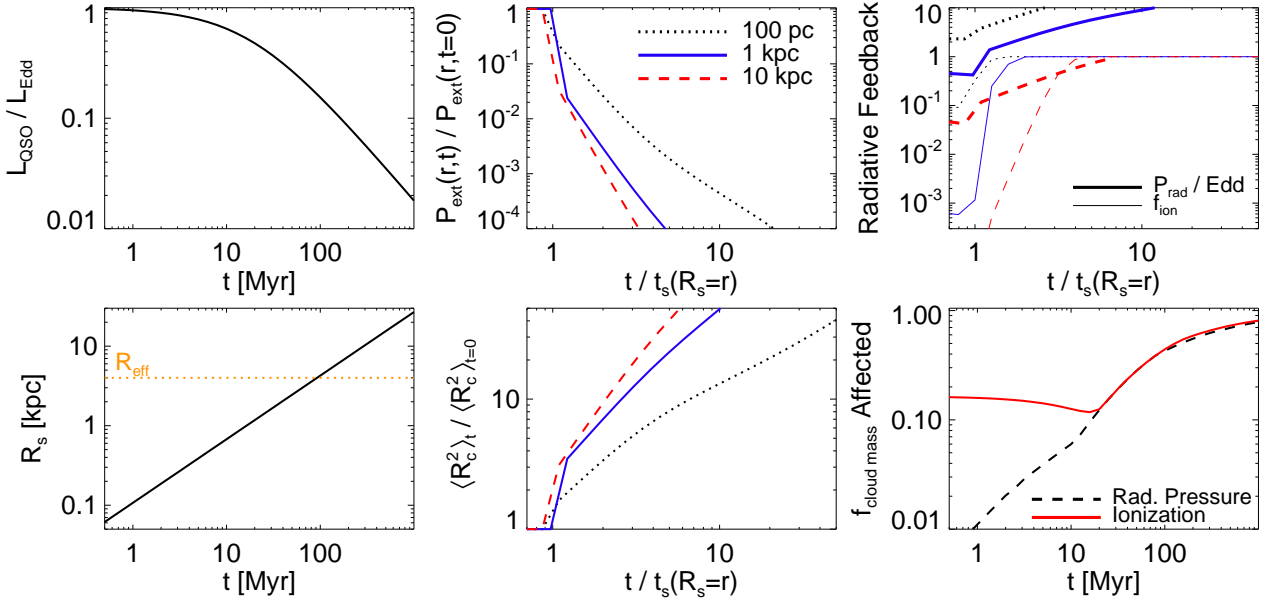}
    \caption{Effects of a two-stage feedback model 
    in the wake of a quasar-induced hot outflow, 
    with parameters of a typical $\sim L_{\ast}$ quasar. {\em Top Left:} 
    Quasar Eddington ratio as a function of time, after feedback from the 
    quasar begins to drive an outflow in the diffuse ISM (time $t=0$). 
    {\em Bottom Left:} Radius of the hot/diffuse outflow as a function of time 
    (compare the galaxy effective radius $R_{\rm eff}$). 
    {\em Top Center:} Change in pressure of the (post-shock) diffuse ISM at a given radius 
    $r$ from the BH, at time $t/t_{s}$ ($t_{s}$ is the time when the hot outflow 
    first reaches $r$). We show three different radii: $r=100\,$pc, $1\,$kpc, 
    and $10\,$kpc. {\em Bottom Center:} Change in effective 
    cross-section of a typical ($R_{0}\sim1\,$pc) cloud at each $r$ 
    versus time $t/t_{s}$. {\em Top Right:} Ratio of 
    the radiation pressure force to the Eddington limit for expulsion of 
    each cloud (thick; $P_{\rm rad}/{\rm Edd}$), and fraction of the 
    cloud ionized (thin; $f_{\rm ion}$). 
    {\em Bottom Right:} Total fraction of cloud mass that can be expelled 
    by radiation pressure or ionized (we consider each separately), 
    as a function of time. The curves reflect the integral over the observed 
    mass spectrum of clouds at each radius, integrated over all 
    radii following the galaxy density profile.  
    \label{fig:clouds}}
\end{figure*}

Figure~\ref{fig:clouds} illustrates some of the basic behaviors 
of this outflow. The energy injection leads to a shock 
that expands outwards with radius $R_{s}(t)$, in an 
approximately power-law fashion as $R_{s}\propto t^{\alpha}$, 
where $\alpha$ is a function of the local density profile, gas 
equation of state, 
and is coupled (weakly) to the declining energy injection rate 
of the BH (Figure~\ref{fig:clouds}; {\em bottom left}). 
For typical galaxy density profiles, a $\gamma=5/3$ gas,
and the conditions 
assumed here, \citet{hopkins:seyferts} show $\alpha\approx4/5$. 
In the wake of the expanding bubble/shock, the post-shock density 
drops in a related power-law manner. Since the spherical accretion rate onto a 
BH scales roughly $\propto \rho$ (in Bondi-Hoyle 
accretion; or similarly $\propto \Sigma$ for viscous accretion from a disk), 
the accretion rate, and hence luminosity $L=0.1\,\dot{M}_{\rm BH}\,c^{2}$, 
will decay as well (Figure~\ref{fig:clouds}; {\em top left}). Again, we refer to the full derivation in 
\citet{hopkins:seyferts} for details, but the self-consistent 
solution derived therein can be approximated as 
$L\propto [1+t/t_{Q}]^{-\eta_{L}}$, with $\eta_{L}\sim1.5$ and 
$t_{Q}\sim1-5\times10^{7}\,$yr for the 
parameters here \citep[consistent with various observational constraints; see][]{martini04,
yu:mdot.dist,hopkins:mdot.dist}. 
The value of $\eta_{L}$ follows from $\alpha$ and 
generic behavior of Sedov-Taylor post-shock gas, and 
$t_{Q}$ is simply related, modulo appropriate numerical coefficients, 
to the local dynamical time near the BH radius of influence. 

The shock therefore crosses a radius $r$ at a time $t_{s}$ (when 
$R_{s}=r$). In the wake of the shock, the post-shock ambient pressure 
$P_{\rm ext}$ will drop, reflecting the density decline from material 
being blown out (Figure~\ref{fig:clouds}; {\em top center}). 
Again, this approximately follows a standard 
decline in the wake of a 
Sedov-Taylor blastwave; the exact solution for the 
pressure internal to the blastwave under the conditions here must be obtained numerically, 
but \citet{ostrikermckee:blastwaves} show that it can be approximated 
as a double power-law. Roughly speaking, there is a rapid drop in pressure in the 
immediate post-shock region, as the thin shell at the front of the 
blastwave clears the diffuse material away from the region, and the pressure declines as a steep power 
$P\propto ([t - t_{s}]/t_{c})^{-\beta}$, where $t_{c}$ is the crossing time of the 
shock relative to the cloud ($\sim R_{c}/v_{s}$) and 
$\beta\sim3-5$ (the exact index depends on the local density profile slope and 
rate of decay of the driving source, so is not the same at all radii). This is followed by a more gradual decline, 
as the diffuse medium internal to the shock relaxes, is heated, and expands, 
with $P\propto ([t-t_{s}]/t_{s})^{-\beta^{\prime}}$ and $\beta^{\prime}\sim 2$ 
\citep[again, for the detailed derivation of the double power-law structure in 
the wake of the blastwave, for the conditions of the feedback-driven 
blastwaves considered here, we refer to][]{hopkins:seyferts}. 

In the wake of the shock, with a rapidly declining background 
pressure and density, the cloud will be mixed and effectively increase its 
surface area. We can solve for the behavior of each cloud -- at least the key parameter of 
interest, the effective radius of the cloud in the direction perpendicular to the shock 
($R_{c}$) as a function of time, according to the approximations in 
\citet{kmc:shock.cloud} in the wake of the shock defined above. 
If the cloud could somehow resist shredding (say, via sufficient magnetic field or turbulent 
support) this would be trivial: the cloud (initial pressure equilibrium) 
cloud would expand isothermally (e.g.\ conserving 
total magnetic energy) as it is now over-pressurized, such that pressure equilibrium 
would be conserved. For a background pressure declining with some power law 
$P\propto (t/t_{s})^{-\beta}$, this implies expansion of the cloud with 
$R_{c}/R_{0} = (t/t_{s})^{\beta/3}$ (generally, $R_{c}/R_{0} = (P_{c}/P_{0})^{-1/3}$, 
for isothermal expansion). If the cloud were supported by thermal energy with 
no new inputs this index would be slightly modified (for adiabatic expansion, 
$R_{c}/R_{0} = (P_{c}/P_{0})^{1/5}$) but the behavior is qualitatively similar. 
Technically this approximation assumes that the time for the cloud to equilibrate 
is short relative to the timescale on which the background is changing, but this is 
easily satisfied. The cloud expands/equilibrates on its internal crossing time, 
given by the effective sound speed as $\sim R_{c}/c_{s,\, \rm eff}$; 
for quasi-virial clouds this is simply the dynamical time $1/\sqrt{G\,\rho}$ 
where the effective density $\rho$ follows from the observed size-mass 
relation (Equation~\ref{eqn:cloud.sizemass}). Using the observed values, 
this gives a timescale of $\sim 0.1-1\times10^{6}\,$yr (for cloud sizes 
$\sim0.1-10\,$pc). Compare this to the characteristic timescale for the 
evolution of the background, a few $t_{s}$ (itself of order $t_{\rm dyn}$, the galaxy dynamical 
time at $R$). For a typical $\sigma \sim 200\,{\rm km\,s^{-1}}$ spheroid, 
$t_{s}\gg R_{c}/c_{s,\, \rm eff}$ for all radii $\gtrsim100\,$pc -- in other words, 
this condition is easily satisfied at the radii $\sim R_{e}$, which contain most of the 
mass of the galaxy. Figure~\ref{fig:clouds} ({\em bottom center}) shows how 
the clouds will expand in effective cross section ($\sim R_{c}^{2}$), relative 
to their initial sizes, given this declining background pressure for the simple isothermal 
case. 

For the more complex case of cloud shredding, it turns out that, 
in aggregate, a similar scaling obtains. 
The characteristic time for the cloud to effectively be mixed via instabilities and 
so effectively increase its cross section is a few cloud-crushing timescales 
$t_{cc} = \chi^{1/2}\,R_{0}/v_{s}$. But since the shock velocity is of order the galaxy 
escape velocity for the ``interesting'' diffuse outflows considered here 
($v_{s}\sim$ a few $\sigma$), and typical $\sigma\sim200\,{\rm km\,s^{-1}}$, 
compared with a typical effective sound speed of a virialized cloud 
$\sim1-10\,{\rm km\,s^{-1}}$, this time is almost always much shorter than 
(or at least comparable to) the cloud dynamical time. 
Following \citet{kmc:shock.cloud}, cloud shredding will equilibrate when 
the system expands by a factor $\sim \chi^{1/2}$ in the perpendicular 
direction (and a small, $\sim$ constant factor in the parallel direction), where 
$\chi$ is the initial density contrast; in other words, until the {\em effective} 
density and pressure drop to approximate equilibrium with the background. 
Thus, for timescales $\sim t_{s}$ over which the background is evolving, 
long compared to the cloud-crushing time, we can consider the systems to 
effectively expand with an average effective radius scaling 
in the same way in equilibrium with the external/hot medium background pressure 
(i.e.\ similar effective net expansion, averaged over these timescales, 
as in the isothermal expansion case).

We then solve for the behavior of each cloud in the wake of this 
hot outflow, according to the approximations in 
\citet{kmc:shock.cloud} and \S~\ref{sec:radiative} (Figure~\ref{fig:clouds}; {\em top right}). 
In particular, given the time-evolution in $R_{c}/R_{0}$ 
shown in Figure~\ref{fig:clouds}, we use the scalings derived in \S~\ref{sec:radiative} 
to estimate the fraction of the cloud 
(at some initial radius $r$) which will be ionized (i.e.\ 
$f_{\rm ion}=h_{\rm ionized}/R_{c}$ from Equation~\ref{eqn:hionized}; 
where the AGN accretion rate $\mdot$ and cloud expansion $R_{c}/R_{0}$ 
are given as a function of time above, the initial cloud radius $r$ is one of those 
specified in the Figure, and we chose a representative initial cloud with 
radius $R_{0,\, \rm pc}=1$ for illustrative purposes). 
We also show the relative strength of radiation pressure on the cloud, 
i.e.\ the radiation pressure relative to the local Eddington limit (that which would 
unbind the cloud), $P_{\rm rad}/{\rm Edd} = 0.14\,\mdot\,(R_{c}/R_{0})^{2}\,[(r+R_{e})/r]^{2}$ 
(re-arranging Equations~\ref{eqn:prad.vs.pedd} \&\ \ref{eqn:rcrit.prad}; where 
again $\mdot$, $R_{c}/R_{0}$, and $r$ for the clouds is given). 
Since these both scale steeply with the increasing effective cloud size 
($\propto R_{c}^{5}$ and $R_{c}^{2}$, respectively), both increase rapidly in time. 
At early times, only the clouds within a narrow region $\sim100\,$pc around the quasar 
are efficiently ionized, and the effects of radiation pressure are weak -- 
similar to what is observed in the narrow-line regions of AGN 
\citep{crenshaw:nlr,rice:nlr.kinematics}. 
This changes rapidly in the wake of the outflow at large radii -- 
the deformation induced makes the clouds vulnerable to ionization and radiative 
momentum driving. 

Integrating over the entire cloud population and galaxy mass, 
i.e.\ integrating over the initial cloud mass or size ($R_{0}$) spectrum at each 
galactic radius $r$, and then over the total cold gas density at each radius $r$, 
we obtain the {\em total} fraction of cloud mass that can be ionized 
or effectively accelerated by radiation pressure (we define the latter 
as the integral of mass in clouds where $P_{\rm rad}/{\rm Edd}>1$). 
Figure~\ref{fig:clouds} ({\em bottom right}) shows this as a function of time 
given the above outflow properties. 
Most of the cold mass can be effectively accelerated by radiation pressure 
at large radii in a timescale $\sim$a few galaxy dynamical times, 
despite the declining AGN luminosity.
We have experimented with varying the 
exact parameters adopted here, and find that the qualitative results are robust. 
In a couple of dynamical times, $\sim90\%$ of the original cold cloud mass 
becomes vulnerable to secondary radiative feedback; i.e.\ these numbers -- the fraction 
that can be ionized and/or the strength of radiation pressure approach or 
exceed unity. 

This and the previous results also provide an important check of an 
implicit assumption in this model: that the cloud acceleration time 
is long relative to the timescale for the clouds to expand/equilibrate. 
If this were not true, the two behaviors could not be treated independently, and 
the physical consequences are unclear (it is possible, for example, that each ``parcel'' of 
the cloud which is mixed or stripped off by instabilities would rapidly be accelerated, leading to 
the edges of the clouds being effectively blown away or stripped but giving little acceleration 
to the cloud core). As noted above, the expansion/equilibration time is 
given by the cloud-crushing time, comparable to the internal 
cloud crossing/dynamical times $<10^{6}\,$yr. In comparison, the acceleration times 
are of order a couple to a few $t_{s}$, the dynamical time at $r$ in the galaxy, 
which (as we discuss above) are generically much larger 
than the cloud crossing time at all radii $\sim R_{e}$ (where most of the galaxy 
mass is located), indeed all radii $\gtrsim100\,$pc (i.e.\ all radii which are not 
already affected by feedback even without a diffuse outflow). In a global sense, 
most of the mass is accelerated and/or ionized over a timescale 
$\sim 10^{7}-10^{8}\,$yr, much longer than the crossing/collapse times of 
all but the most massive molecular cloud complexes. 

In fact, this acceleration time 
is a relatively long time, at large scales ($\sim$a few $10^{8}\,$yr), 
and the AGN luminosity has correspondingly decayed to $\sim 1\%$ of the 
Eddington limit. The model above accounts for this, but an important 
question remains if real AGN can sustain even this level of energetic output 
over these time intervals. If, for example, 
AGN switch to a radiatively inefficient state above or around this 
accretion rate, the driving will suddenly vanish. Clearly, this is 
an interesting regime; better knowledge of how feedback-induced 
hot outflows and subsequent lightcurve evolution proceed 
will be important to understanding both how dramatic the effect on the 
galaxy will be and, potentially, how much variation there may be 
between galaxies.

\section{Discussion}
\label{sec:discuss}

``Feedback'' from bright AGN is a topic of 
fundamental interest for galaxy evolution, but it remains 
unknown whether or not any of the obvious candidate feedback mechanisms 
are capable of effectively coupling to cold molecular gas, especially at kpc scales,
the dominant reservoir for star formation. Here, we demonstrate 
that it is at least possible that the cold gas reservoir is destroyed and/or blown out 
of the galaxy 
{\em despite} inefficient coupling of ``initial'' feedback mechanisms that originate 
near the BH. 

If some coupling of energy or momentum near the BH -- whether from 
e.g.\ Compton heating, radiation pressure, BAL winds, jets, or resonant line-driving -- 
can generate a wind or shock/blastwave in the warm/hot ISM, then when the outflow 
passes by a cold cloud, even if it does not directly entrain the material, it will 
generate various instabilities that ``shred'' the cloud and mix it, efficiently 
enhancing the cloud cross section in the perpendicular direction. Even if the 
cloud is magnetically supported or extremely dense and able to resist instabilities, 
there is still a growing pressure imbalance that drives the cloud 
to expand in the same manner. 

This is well-studied in the context of supernovae-driven winds, but there is 
an important difference 
in the presence of a bright quasar. 
The effective increase in cross section means that momentum driving 
and ionization heating from 
the quasar radiation is quickly able to act in much more dramatic fashion on clouds 
that were once too dense and too small (or at too large a distance from the 
black hole) to be perturbed by the radiation field.
This effect can have dramatic implications for star formation in quasar host galaxies.

Because radiation pressure always acts, this means the energy needed in 
``initial'' feedback from the central source to e.g.\ drive winds in the low-density 
hot gas will be much less than if it were expected to act directly on the 
cold clouds. We show that the energetic or momentum driving requirements for the 
initially driven feedback are reduced by at least factor $f_{\rm hot}\sim0.1$ 
(the mass fraction in the hot diffuse ISM); i.e.\ rather than 
the canonical $\sim5\%$ of the radiant energy ($\sim100\%$ 
momentum) needed in 
the initial outflow if it were to 
entrain the entire gas supply directly, only $\sim0.5\%$ ($\sim10\%$ momentum) is 
sufficient to drive the hot gas and enter the regime of interest here. 
Another way of stating this is, for accretion with an Eddington ratio 
$\mdot$ and BH mass $M_{\rm BH}$ relative to the expectation 
$\langle M_{\rm BH} \rangle$ from the $M_{\rm BH}-\sigma$ relation, 
the relevant outflows will be driven (and star formation 
suppressed) when
\begin{equation}
\eta\,\mdot\,\frac{M_{\rm BH}}{\langle M_{\rm BH} \rangle}\sim 0.05\,f_{\rm hot}\ ,
\end{equation}
where $\eta$ is the feedback efficiency ($\dot{E}=\eta\,L$). 
Given this criterion, that there is sufficient momentum in photons 
for the ``secondary'' feedback to act is gauranteed for all 
but the most extremely gas-dominated systems. 

Note that the derivation here pertains to large clouds 
($R_{0}\gtrsim$\,pc), observed to be in rough pressure equilibrium with 
the ambient medium and containing most of the ISM mass. Dense cores ($R_{0}\ll$\,pc) 
are observed to be in self-gravitating collapse; these will continue to collapse 
and form stars on a very short timescale despite a diffuse outflow. 
The important thing is that no {\em new} cold gas reservoir of large 
clouds will be available to form new cores. 

We have also neglected the possibility that galaxies are 
highly self-shielding. For example, in dense nuclear star-forming regions 
in e.g.\ ULIRGs, the column densities are so high 
\citep[$N_{\rm H}\gtrsim 10^{23}\,{\rm cm^{-2}}$, see e.g.][]{komossa:ngc6240,
li:radiative.transfer} that the quasar can do little until star formation 
exhausts more of the gas supply. In disk-dominated galaxies, gas near 
$R_{\rm eff}$ can be similarly self-shielded. If the radiation is isotropic, 
then for some disk mass and gas fraction, only a fraction $\sim h/R$ 
(the fractional scale height at $R$) will couple to the relevant area (and 
the radiation may, in fact, be preferentially polar, yielding even lower 
efficiency). Only the most gas-poor disks, or the central regions of disks 
(where systems are typically bulge-dominated) will be affected 
by the coupling efficiencies above. 

What we outline here is a simple model for the qualitative physical 
effects that may happen when cold clouds in the ISM encounter a hot outflow 
driven by an AGN. More detailed conclusions 
will require study in hydrodynamic simulations which incorporate gas 
phase structure, cooling, turbulence, self-gravity, radiation transport, 
and possibly (if they provide significant pressure support) magnetic fields. 
Detailed effects which we cannot follow analytically, 
such as e.g.\ self-shielding within thin, dense fingers in 
Rayleigh-Taylor or Kelvin-Helmholtz instabilities may alter the effects of the 
radiation field on the cloud materia and change our conclusions. 
Nevertheless, our simple 
calculations here demonstrate that the process of cloud deformation in the wake 
of a hot outflow can have dramatic implications for the susceptibility of 
those clouds to other modes of feedback, and should motivate 
further study. 

According to these simple considerations, 
outflows driven by AGN feedback may in fact be ``multi-stage'' or ``two-tiered'', with 
an initial hot shockwave or strong wind driven by feedback mechanisms near the 
BH, which is then supplemented by a successive wind driven out as clouds in the 
wake of the former are deformed/mixed and increase their effective 
cross-section to the AGN luminosity. The characteristic velocity of this secondary outflow, 
which will carry most of the mass, should be $\sim v_{\rm esc}$ at 
the radii of launching ($\sim10^{2}\,{\rm km\,s^{-1}}$), and it will behave similarly 
to outflows from star formation. Indeed, because the driving occurs at large radii, 
it is not clear whether it could be distinguished from 
stellar-driven outflows at all, except indirectly (e.g.\ in cases where the observed 
star formation is insufficient to power the outflow). 
Charactersitic timescales are 
$\sim$ a few $t_{\rm dyn}$ of the galaxy, 
so much of the outflow occurs as sub-Eddington luminosities 
(as the AGN fades in the wake of launching the ``primary'' outflow) 
and the systems will not appear gas-depleted until they have evolved by a significant 
amount ($\sim$few $10^{8}$yr, at Eddington ratios $\sim0.01$ typical of ``quiescent'' 
ellipticals). These processes should nevertheless imply effective shutdown of 
star formation and destruction/heating of the cold gas supply 
in ``massive'' BH systems -- bulge-dominated systems on the $M_{\rm BH}-\sigma$ 
relation that have recently been excited to near-Eddington luminosities. 

Most intriguing, this reduces the energetic requirements for the ``initial'' feedback -- 
whatever might drive an outflow in the hot gas from the vicinity of the BH -- 
by an order of magnitude. Our estimates suggest that coupling only a fraction 
$\sim10^{-3}$ of the luminosity of the AGN on small scales would be sufficient 
to drive such a hot outflow and then allow $\sim100\%$ of the 
radiative energy/momentum to couple to cold gas. If, for example, 
quasar accretion-disk (or broad-line) winds (with characteristic velocities 
$v\sim10^{4}\,{\rm km\,s^{-1}}$) do not immediately 
dissipate all their energy, then 
the hot outflows we invoke would be generated with a mass-loading in such 
winds of just $\sim0.1\,\dot{M}_{\rm BH}$, a fraction of the BH accretion rate.

\acknowledgments 
We thank Lars Hernquist and Eliot Quataert for helpful discussions 
in the development of this work, and thank Vincenzo Antonuccio-Delogu, 
Pat Hall, and Barry McKernan for helpful comments on an earlier draft. 
We also appreciate the hospitality of the 
Aspen Center for Physics, where this paper was partially developed. 
Support for PFH was provided by the Miller Institute for Basic Research 
in Science, University of California Berkeley. 
\\

\bibliography{/Users/phopkins/Documents/lars_galaxies/papers/ms}

\end{document}